# Structure and tie strengths in mobile communication networks

J.-P. Onnela[*,†], J. Saramäki[†], J. Hyvönen[†], G. Szabó[‡,§],
D. Lazer[¶], K. Kaski[†], J. Kertész[∥], A.-L. Barabási[‡,§]

[*]Physics Department, Clarendon Laboratory, Oxford University, Oxford, OX1 3PU, U.K.; [†]Laboratory of Computational Engineering, Helsinki University of Technology, P.O.Box 9203, FI-02015 TKK, Finland; [‡]Department of Physics and Center for Complex Networks Research, University of Notre Dame, IN 46556, USA; [§]Center for Cancer Systems Biology, Dana Farber Cancer Institute, Harvard University, Boston, MA 02115, USA; [¶]John F. Kennedy School of Government, Harvard University, Cambridge, MA 02138, USA; and [∥]Department of Theoretical Physics, Budapest University of Technology and Economics, H1111 Budapest, Hungary

[*]To whom correspondence should be addressed. E-mail: jp.onnela@physics.ox.ac.uk.

**Electronic databases, from phone to emails logs (1, 2, 3, 4), currently provide detailed records of human communication patterns, offering novel avenues to map and explore the structure of social and communication networks. Here we examine the communication patterns of millions of mobile phone users, allowing us to simultaneously study the local and the global structure of a society-wide communication network. We observe a coupling between interaction strengths and the network's local structure, with the counterintuitive consequence that social networks are robust to the removal of the strong ties, but fall apart following a phase transition if the weak ties are removed. We show that this coupling significantly slows the diffusion process, resulting in dynamic trapping of information in communities, and find that when it comes to information diffusion, weak and strong ties are both simultaneously ineffective.**

Uncovering the structure and function of communication networks has always been constrained by the practical difficulty of mapping out interactions among a large number of individuals. Indeed, most of our current understanding of communication and social networks (5) is based on questionnaire data, reaching typically a few dozen individuals and relying on the individual's opinion to reveal the nature and the strength of the ties. The fact that currently an increasing fraction of human interactions are recorded from email (1, 2, 3) to phone records (4), offers unprecedented opportunities to uncover and explore the large scale characteristics of communication and social networks. Here we take a first step in this direction by exploiting the widespread use of mobile phones to construct a map of a society-wide communication network, capturing the mobile interaction patterns of millions of individuals. The dataset allows us to explore the relationship between the topology of the network and the tie strengths between individuals, information that was inaccessible at the societal level before. We demonstrate a local coupling between tie strengths and network topology, and show that this coupling has important consequences for the network's global stability if ties are removed, as well as for the spread of news and ideas within the network.

A significant portion of a country's communication network was reconstructed from 18 weeks of all mobile phone call records among approximately 20% of the country's entire population, 90% of whose inhabitants had a mobile phone subscription (see Supporting Information). While a single call between two individuals during 18 weeks may not carry much information, reciprocal calls of long duration between two users serves as a signature of some work, family, leisure or service based relationship. Therefore, in order to translate the phone log data into a network representation that captures the characteristics of the underlying communication network, we connect two users with an undirected link if there has been at least one reciprocated pair of phone calls between them—i.e., A called B, and B called A— and define the strength, $w_{AB} = w_{BA}$, of a tie as the aggregated duration of calls between users A and B. This procedure eliminates a large number of one-way calls, most of which correspond to single events, suggesting that they typically reach individuals that the caller does not know personally. The resulting mobile call graph (MCG) (4) contains $N = 4.6 \times 10^6$ nodes and $L = 7.0 \times 10^6$ links, the vast majority (84.1%) of these nodes belonging to a single connected cluster (giant component, GC). Given the very large number of users and communication events in the database, we find that the statistical characteristics of the network and the GC are largely saturated, observing little difference between a two- or a three-month long sample. Note that the MCG captures only a subset of all interactions between individuals, a detailed mapping of which would require face-to-face, email and landline communications as well. Yet, while mobile phone data capture just a slice of communication among people, research on media multiplexity suggests that the use of one medium for communication between two people implies communication via other means as well (6). Furthermore, in the absence of directory listings, the mobile phone data is skewed towards trusted interactions (that is, people tend to share their mobile numbers only with individuals they trust). Therefore, the MCG can be used as a proxy of the communication network between the users. It is of sufficient detail to allow us to address the large scale features of the underlying human communication network, and the major trends characterizing it.

The MCG has a skewed degree distribution with a fat tail (Fig. 1A), indicating that while most users communicate with only a few individuals, a small minority talk with dozens (4, 7). If the tail is approximated by a power law, the exponent $\gamma_k = 8.4$ is significantly higher than the value observed for land lines ($\gamma = 2.1$ for the in-degree distribution, see Ref. *8*). For such a rapidly decaying degree distribution the hubs are few, and therefore many properties of traditional scale-free networks, from anomalous diffusion (9) to error tolerance (10) are absent. This is probably rooted in the fact that institutional phone numbers, corresponding to the vast majority of large hubs in the case of land lines, are absent, and in contrast with email, where a single email can be sent to many recipients, resulting in well-connected hubs (1), a mobile phone conversation typically represents a one-to-one communication. The tie strength distribution is broad (Fig. 1B), however, decaying with an exponent $\gamma_w = 1.9$, so that while the majority of ties correspond to a few minutes of air time, a small fraction of users spend hours chatting with each other. This is rather unexpected, given that fat tailed tie strength distributions have been observed mainly in networks characterized by global *transport processes*, such as the number of passengers carried by the airline transportation network (11), the reaction fluxes in metabolic networks (12) or packet transfer on the Internet (13), in which case the individual fluxes are determined by the global network topology. An important feature of such global flow processes is local conservation: all passengers arriving to an airport need to be transported away; each molecule created by a reaction needs to be consumed by some other reaction; or each packet arriving to a router needs to be sent to other routers. While the main purpose of the phone is information transfer between two individuals, such local conservation that constrains or drives the tie strengths are largely absent, making any relationship between the topology of the MCG and local tie strengths less than obvious.

Complex networks often organize themselves according to a global efficiency principle, meaning that the tie strengths are optimized to maximize the overall flow in the network (13, 14). In this case the weight of a link should correlate with its betweenness centrality, which is proportional to the number of shortest paths between all pairs of nodes passing through it (13, 15, 16, 17). Another possiblility is that the strength of a particular tie depends only on the nature of the relationship between two individuals, and is thus independent of the network surrounding the tie (dyadic hypothesis). Finally, the much studied strength of weak ties hypothesis (18, 19, 20) states that the strength of a tie between A and B increases with the overlap of their friendship circles, resulting in the importance of weak ties in connecting communities. This leads to high betweenness centrality for weak links, which can be seen as the mirror image of the global efficiency principle.

In Fig. 2A we show the network in the vicinity of a randomly selected individual, where the link color corresponds to the strength of each tie. It appears from this figure that the network consists of small local clusters, typically grouped around a high degree individual. Consistent with the strength of weak ties hypothesis, the majority of the strong ties are found within the clusters, indicating that users spend most of their on-air time talking to members of their immediate circle of friends. In contrast, most links connecting different communities are visibly weaker than the links within the communities. As a point of comparison, when we randomly permute the link strengths among the connected user pairs (Fig. 2B), in what would be consistent with the dyadic hypothesis, we observe dramatically more weak ties within the communities and more strong ties connecting distinct communities. Finally, even more divergent with the observed data (Fig. 2A), we illustrate what the world would be like if, as predicted by the global efficiency principle and betweenness centrality, inter-community ties ("bridges") were strong and intra-community ties ("local roads") weak (Fig. 2C). In order to quantify the differences observed in Fig. 2 we measure the relative topological overlap of the neighborhood of two users $v_i$ and $v_j$, representing the proportion of their common friends (21) $O_{ij} = n_{ij}/((k_i - 1) + (k_j - 1) - n_{ij})$, where $n_{ij}$ is the number of common neighbors of $v_i$ and $v_j$, and $k_i$ ($k_j$) denotes the degree of node $v_i$ ($v_j$). If $v_i$ and $v_j$ have no common acquaintances we have $O_{ij}=0$, the link between *i* and *j* representing potential bridges between two different communities. If *i* and *j* are part of the same circle of friends then $O_{ij}=1$ (Fig. 1C). The dyadic hypothesis implies the absence of a relationship between the local network topology and weights, and, indeed, we find that permuting randomly the tie strengths between the links results in $O_{ij}$ that is independent of $w_{ij}$ (Fig. 1D). We find, however, that according to the global efficiency principle $\langle O \rangle_b$ decreases with the betweenness centrality $b_{ij}$, indicating that on average the links with the highest betweenness centrality $b_{ij}$ have the smallest overlap. In contrast, for the real communication network $\langle O \rangle_w$ increases as a function of the percentage of links with weights smaller than *w*, demonstrating that the stronger the tie between two users, the more their friends overlap, a correlation that is valid for approximately 95% of the links (Fig. 2D). This result is broadly consistent with the strength of weak ties hypothesis, offering its first societal-level confirmation. It suggests that tie strength is, in part, driven by the network structure in the tie's immediate vicinity. This is in contrast with a purely dyadic view, according to which the tie strength is determined only by the characteristics of the individuals it connects, or the global view, which asserts that tie strength is driven by the whole network topology.

    To understand the systemic or global implications of this local relationship between tie strength and network structure, we explore the network's ability to withstand the removal of either strong or weak ties. To evaluate the impact of removing ties, we measure the relative size of the giant component $R_{GC}(f)$, providing the fraction of nodes that can all reach each other through connected paths as a function of the fraction of removed links, *f*. We find that removing in rank order the weakest (or smallest overlap) (Fig. 3, A and B) to strongest (greatest overlap) ties leads to the network's sudden disintegration at $f^w=0.8$ ($f^O=0.6$). In contrast, removing first the strongest (or highest overlap) (Fig. 3, A and B) ties will shrink the network, but will not precipitously break it apart. The precise point at which the network disintegrates can be determined by monitoring $\tilde{S}=\sum_{s<s_{max}} n_s s^2 / N$, where $n_s$ is the number of clusters containing *s* nodes. According to percolation theory, if the network collapses via a phase transition at $f_c$, then $\tilde{S}$ diverges as *f* approaches $f_c$ (22,23). Indeed, we find that $\tilde{S}$ develops a peak if we start with the weakest (or smallest overlap) links (Fig. 3, C and D).

Finite size scaling, a well-established technique for identifying the phase transition, indicates that the values of the critical points are $f_c^O(\infty) = 0.62 \pm 0.05$ and $f_c^w(\infty) = 0.80 \pm 0.04$ for the removal of the weak ties, but there is no phase transition when the strong ties are removed first.

Taken together, these results document a fundamental difference between the global role of the strong and weak ties in social networks: the removal of the weak ties leads to a sudden, phase transition driven collapse of the whole network. In contrast, the removal of the strong ties results only in the network's gradual shrinking, but not its collapse. This is somewhat unexpected, since in most technological and biological networks the strong ties are believed to play a more important structural role than the weak ties, and in such systems the removal of the strong ties leads to the network's collapse (10, 24, 25). This counterintuitive finding underlies the distinct role weak and strong ties play in a social network: given that the strong ties are predominantly within the communities, their removal will only locally disintegrate a community but not affect the network's overall integrity. In contrast, the removal of the weak links will delete the bridges that connect different communities, leading to a phase transition driven network collapse.

The purpose of the mobile phone is information transfer between two individuals. Yet, given that the individuals are embedded in a social network, mobile phones allow news and rumors to diffuse beyond the dyad, occasionally reaching a large number of individuals, a much studied diffusion problem in both sociology (26) and network science (7). Yet, most of our current knowledge about information diffusion is based on analyses of unweighted networks, where all tie strengths are considered equal (26). To see if the observed local relationship between the network topology and tie strength affects global information diffusion, at time zero we infect a randomly selected individual with some novel information. We assume that at each time step each infected individual $v_i$ can pass the information to its contact $v_j$ with effective probability $P_{ij} = xw_{ij}$, where the parameter $x$ controls the overall spreading rate (see Supplementary Information). Therefore, the more time two individuals spend on the phone, the higher the chance that they will pass on the monitored information. As a control, we consider spreading on the same network, but replace all tie strengths with their average value $\overline{w}$, resulting in a constant transmission probability for all links.

As Fig. 4A shows (the real diffusion simulation), we find that information transfer is significantly faster on the network for which all weights are equal, the difference rooted in a dynamic trapping of information in communities. Such trapping is clearly visible if we monitor the number of infected individuals in the early stages of the diffusion process (Fig. 4B). Indeed, we observe rapid diffusion within a single community, corresponding to fast increases in the number of infected users, followed by plateaus, corresponding to time intervals during which no new nodes are infected before the news escapes the community. When we replace all link weights with an average value $\overline{w}$ (the control diffusion simulation) the bridges between communities are strengthened, and the spreading becomes a predominantly global process, rapidly reaching all nodes through a hierarchy of hubs (24).

The dramatic difference between the real and the control spreading process raises an important question: where do individuals get their information? We find that the distribution of the tie strengths through which each individual was first infected (Fig. 4C) has a prominent peak at $w \sim 10^2$ seconds, indicating that in the vast majority of cases an individual learns about the news through ties of intermediate strength. The distribution changes dramatically in the control case, however, when all tie strength are taken to be equal during the spreading process. In this case the majority of infections take place along the ties that are otherwise weak (Fig. 4D). Therefore, in contrast with the celebrated role of weak ties in information access (18, 20), we find that both weak and strong ties have a relatively insignificant role as conduits for information ("the weakness of weak *and* strong ties"): the former because the small amount of on-air time offers little chance of information transfer, and the latter because they are mostly confined within communities, with little access to new information.

To illustrate the difference between the real and the control simulation, we show the spread of information in a small neighborhood (Fig. 4, E and F). First, the overall direction of information flow is systematically different in the two cases, as indicated by the large shaded arrows. In the control runs the information mainly follows the shortest paths. When the weights are taken into account, however, information flows along a strong tie backbone, and large regions of the network, connected to the rest of the network via weak ties, are only rarely infected. For example, the lower half of the network is rarely infected in the real simulation but always infected in the control run. Therefore, the diffusion mechanism in the network is drastically altered when we neglect the tie strengths, responsible for the differences between the curves seen in Fig. 4A and 4B.

While the study of communication and social networks has a long history, examining the relationship between tie strengths and topology in society-spanning networks has generally been impossible. In this paper, taking advantage of society-wide data collection capabilities offered by mobile phone logs, we show that tie strengths correlate with the local network structure around the tie, and both the dyadic hypothesis and the global efficiency principle are unable to account for the empirical observations.

It has been long known that many networks show resilience to random node removal, but are fragile to the removal of the hubs (10, 27, 28, 29). In terms of the links, one would also expect that the strong ties play a more important role in maintaining the network's integrity than the weak ones. Our analyses document the opposite effect in communication networks: the removal of the weak ties results in a phase transition like network collapse, while the removal of strong ties has little impact on the network's overall integrity. Furthermore, we find that the observed coupling between the network structure and tie strengths significantly slows information flow, trapping it in communities, explaining why successful searches in social networks are conducted primarily through intermediate to weak strength ties while avoiding the hubs (3). Therefore, to enhance the spreading of information, one needs to intentionally force it through the weak links or, alternatively, adopt an active information search procedure.

Taken together, weak ties appear to be crucial for maintaining the network's structural integrity, but strong ties play an important role in maintaining local communities. Both weak and strong ties are ineffective, however, when it comes to information transfer, given that most news in the real simulations reaches an individual for the fist time via ties of intermediate strength.

The observed coupling between tie strengths and local topology has significant implications for our ability to model processes taking place in social networks. Indeed, many current network models either assign the same strength to all ties, or assume that tie strengths are determined by the network's global characteristics, such as betweenness centrality. In addition, some of the most widely used algorithms used to identify communities and groups in complex networks use either betweenness centrality (16), or are based on topological measures (30). Our finding that link weights and betweenness centrality are negatively correlated in mobile communication networks, together with the insights provided by the visually apparent community structure (Fig. 2), offer new opportunities to design  clustering algorithms that are tailored to communication networks, and force us to reevaluate many results that were obtained on unweighted graphs. Putting the structural and functional pieces together, we conjecture that communication networks are better suited to local information processing than global information transfer, a result that has the makings of a paradox. Indeed, the underlying reason for characterizing communication networks with global network concepts, such as path length and betweenness centrality, is rooted in the expectation of communication networks to transmit information globally.

## Acknowledgements

The authors would like to thank Tamás Vicsek for useful discussions. JPO is grateful to the ComMIT Graduate School, and Finnish Academy of Science and Letters, Väisälä Foundation, for a travel grant to visit A-L Barabási at Harvard University. This research was partially supported by the Academy of Finland, Centres of Excellenc programmes, project no. 44897 and 213470 and OTKA K60456. GS and ALB were supported by the NSF-ITR, NSF-DDAAS projects and by the McDonell Foundation.

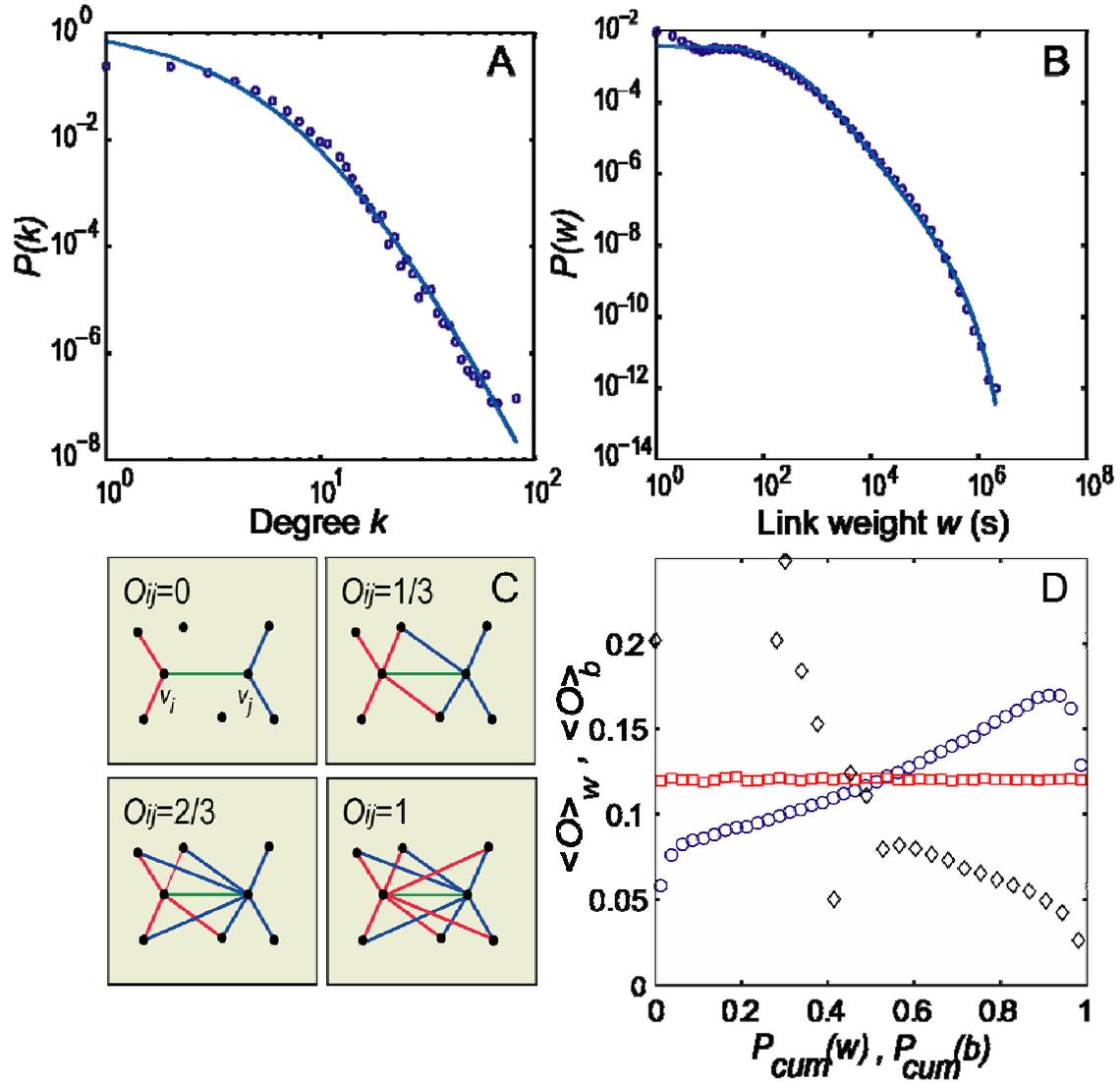

**Fig. 1.** Characterizing the large scale structure and the ties strengths of the mobile call graph. **(A)** Vertex degree and **(B)** tie strength distribution. Each distribution was fitted with $P(x) = a(x + x_0)^{-\gamma} \exp(-x/x_x)$, shown as a blue curve, where $x$ corresponds to either $k$ or $w$. The parameter values for the fits are $k_0 = 10.9$, $\gamma_k = 8.4$, $k_x = \infty$ (A, degree), and $w_0 = 280$, $\gamma_w = 1.9$, $w_x = 3.45 \times 10^5$ (B, weight). **(C)** Illustration of the relative neighborhood overlap between two nodes $v_i$ and $v_j$, its value being shown for four local network configurations. **(D)** In the real network the overlap $\langle O \rangle_w$ (blue o) increases as a function of cumulative tie strength $P_{cum}(w)$, representing the fraction of links with tie strength smaller than $w$. The dyadic hypothesis is tested by randomly permuting the weights, which removes the coupling between $<O>_w$ and $w$ (red □). The overlap $\langle O \rangle_b$ decreases as a function of cumulative link betweenness centrality $b$ (black ♦).

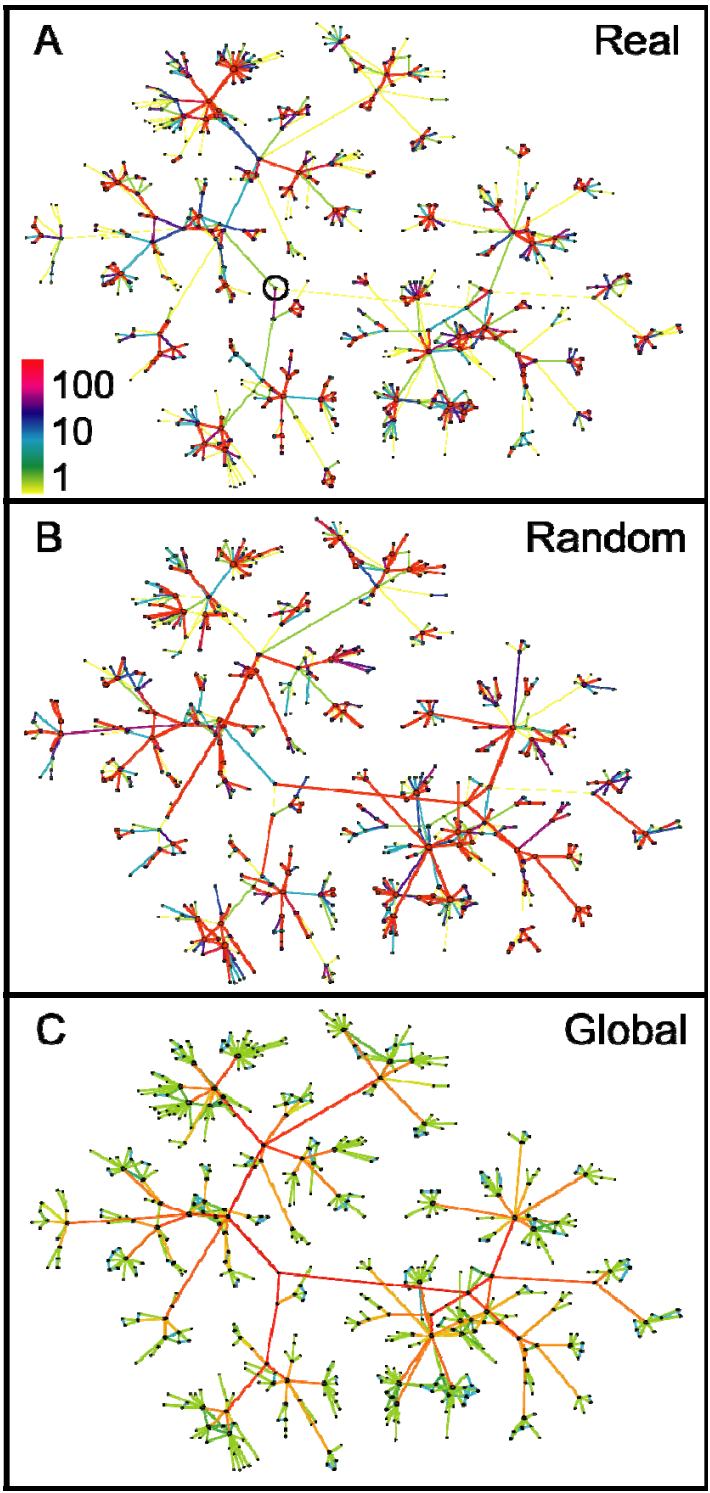

**Fig. 2.** The structure of the mobile call graph around a randomly chosen individual. Each link represents mutual calls between the two users, and all nodes are shown that are at distance less than six from the selected user, marked by a circle in the center. **(A)** The real tie strengths, observed in the call logs, defined as the aggregate call duration in minutes (see color bar). **(B)** The dyadic hypothesis suggests that the tie strength depends only on the relationship between the two individuals. To illustrate the tie strength distribution in this case, we randomly permuted tie strengths for the sample in (A). **(C)** The weight of the links assigned based on their betweenness centrality $b_{ij}$ values for the sample in (A), as suggested by the global efficiency principle. In this case the links connecting communities have high $b_{ij}$ values (red), whereas the links within the communities have low $b_{ij}$ values (green).

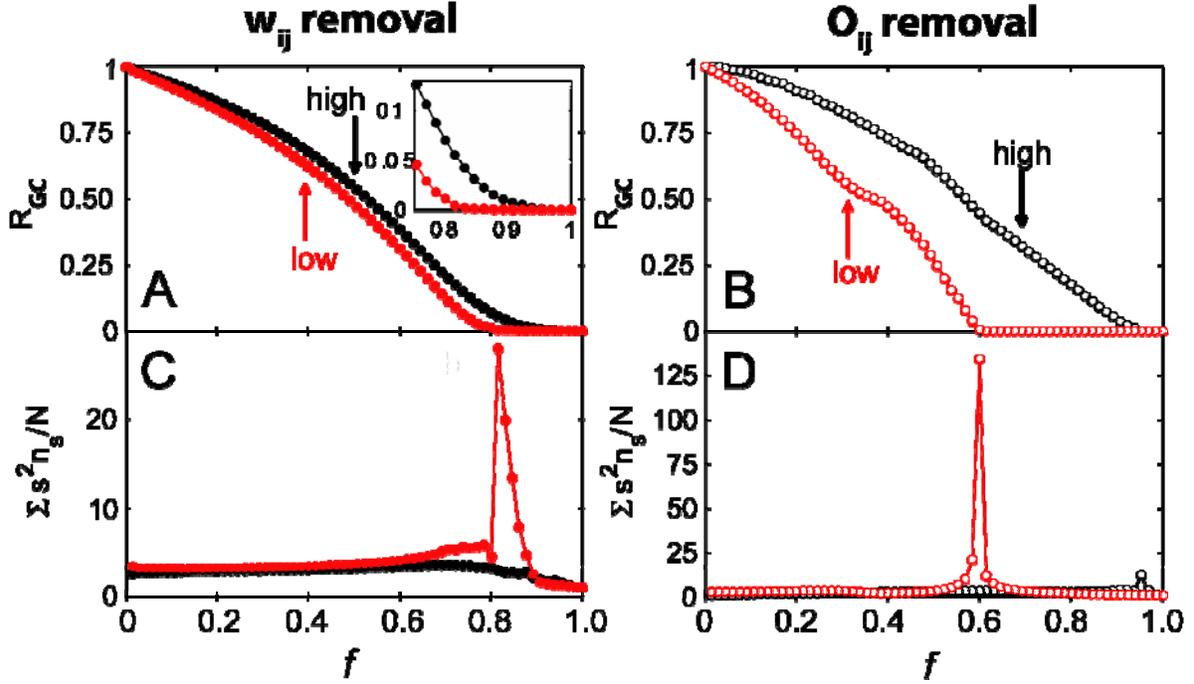

**Fig. 3.** The stability of the mobile communication network to link removal. The control parameter $f$ denotes the fraction of removed links. The panels on the left correspond to the case when the links are removed based on their strengths while on the right panels links were removed based on their overlap. The black curves correspond to removing first the high tie strength (or high $O_{ij}$) links, moving towards the weaker ones, while the red curves represent the opposite, starting with the low strength (or low $O_{ij}$) ties, and moving towards the stronger ones. (**A**, **B**) The relative size of the largest component $R_{GC}(f) = N_{GC}(f)/N_{GC}(f=0)$ indicates that the removal of the low $w_{ij}$ or $O_{ij}$ links leads to a breakdown of the network, while the removal of the high $w_{ij}$ or $O_{ij}$ links leads only to the network's gradual shrinkage. The inset in A shows the blowup of the high $w_{ij}$ region, indicating that when the low $w_{ij}$ ties are removed first, the red curve goes to zero at a finite $f$ value. (**C**, **D**) According to percolation theory, $\tilde{S} = \sum_{s<s_{max}} n_s s^2 / N$ diverges for $N \to \infty$ as we approach the critical threshold $f_c$, where the network falls apart. If we start link removal from links with low $w_{ij}$ (C) or $O_{ij}$ (D) values, we observe a clear signature of divergence. In contrast, if we start with high $w_{ij}$ (C) or $O_{ij}$ (D) links there the divergence is absent. Finite size scaling shows that the small local maximum seen in D at $f \approx 0.95$ does not correspond to a real phase transition (see Supplementary Information).

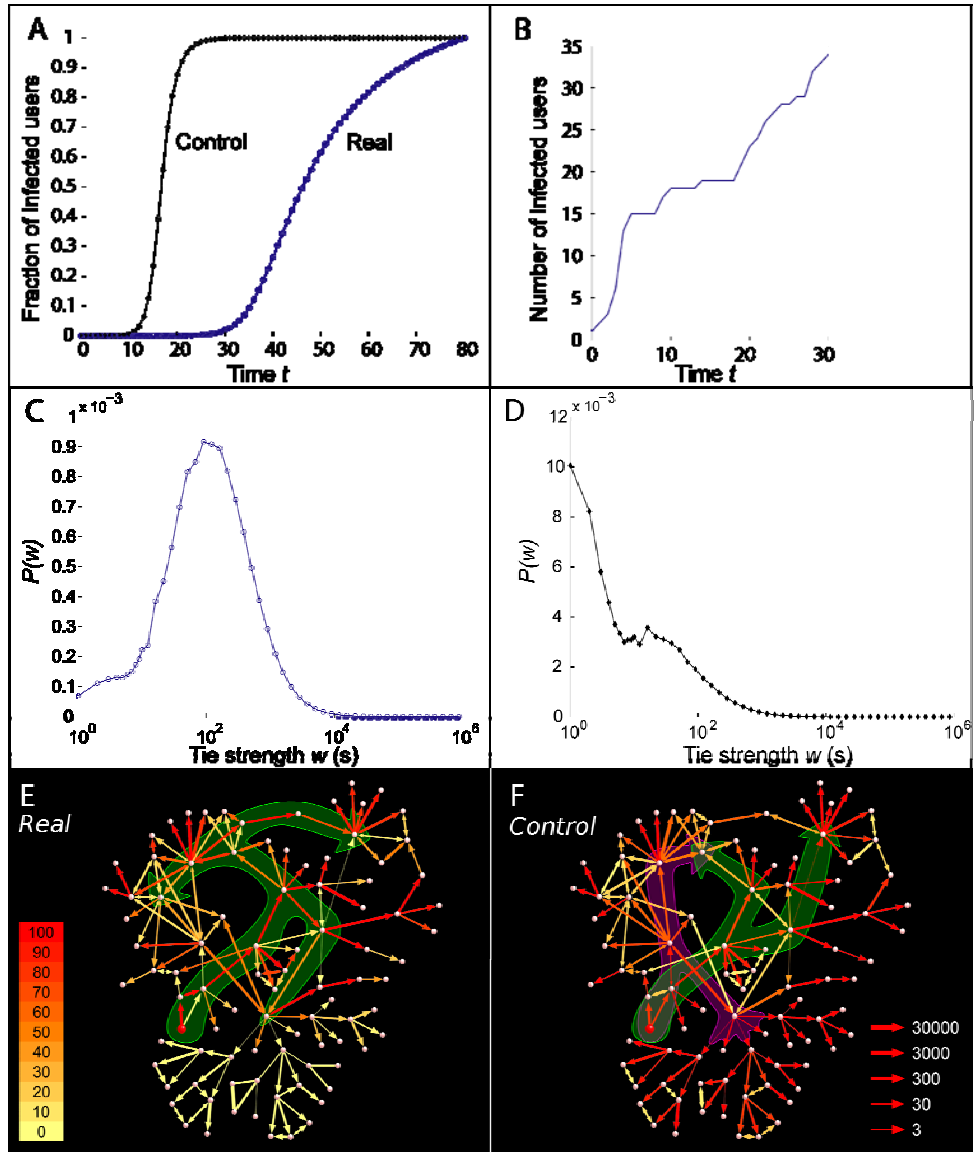

**Fig. 4.** The dynamics of spreading on the weighted mobile call graph, assuming that the probability for a node $v_i$ to pass on the information to its neighbor $v_j$ in one time step is given by $P_{ij} = x w_{ij}$ with $x = 2.59 \times 10^{-4}$. **(A)** The fraction of infected nodes as a function of time $t$. The blue curve (o) corresponds to spreading on the network with the real tie strengths, while the black curve (∗) represents the control simulation, where all tie strengths are considered equal. **(B)** Number of infected nodes as a function of time for a single realization of the spreading process. Each steep part of the curve corresponds to invading a small community. The flatter part indicates that the spreading becomes trapped within the community. Distribution of strengths of the links responsible for the first infection for a node in the **(C)** real network and **(D)** control simulation. **(E,F)** Spreading in a small neighborhood in the simulation using the real (E) weights or the control case, when all weights are taken to be equal (F). The infection in all cases was released from the node marked in red, and the empirically observed tie strength is shown as the thickness of the arrows (right scale). The simulation was repeated 1000 times; the size of the arrowheads is proportional to the number of times that information was passed in the given direction, and the color indicates the total number of transmissions on that link (the numbers in the color scale refer to percentages of 1000). The contours are guide to the eye, illustrating the difference in the information direction flow in the two simulations.

**SUPPLEMENTARY INFORMATION**

**Preparing the data**

The data used in this study was obtained from a mobile phone operator, from now on referred to as the "operator". We focus exclusively on voice calls, filtering out all other services, such as voice mail, data calls, text messages, chat, and operator calls. For the purpose of retaining customer anonymity, each subscription is identified by a surrogate key such that it is not possible to recover the actual phone numbers from it. Since there is no other information available for identifying or locating customers, this guarantees that their privacy is respected. We have filtered out calls that involve other operators, incoming or outgoing, keeping only those transactions in which the calling and receiving subscription is governed by the operator. This filtering is needed to eliminate the bias between the operator and other mobile service providers as we have a full access to the customers of the operator, but only partial access to the activity of other providers.

A small fraction of the subscriptions appears to be used for business or business-like purposes, which appear as users with a very large number of calls never returned. To ensure that we are dealing with genuine social interactions, we require links to represent reciprocal calls within the investigated time period, so that *A* needs to call *B* and vice versa for a link to be placed between them. This restriction eliminates telemarketing calls and wrong numbers. It is possible that this induces some false negatives, i.e. some links corresponding to genuine social interaction may go undetected. However, since the monitored time window is relatively long, over one third of a year, there is plenty of time to reciprocate the calls, limiting the number of false negatives.

Two quantities could be used as tie strengths: the total number and the total duration of calls placed within the period. As expected, these two variables are statistically dependent, giving rise to Pearson's linear correlation coefficient of 0.70. We have chosen to use call durations as weights (or tie strengths) $w_{ij}$, since they implicate the temporal and financial commitment (billing is based on call duration) to the relationship. In addition, since call durations are measured in seconds, they can be considered a continuous weight variable, whereas the number of calls suffers from strong discretization.

Given the way we have constructed the network, an interaction, or link, corresponds to a *social association* between two individuals and it is by nature bi-directional. It would be possible to retain directions in the network using directed links and thus have asymmetric weights, i.e. $w_{ij} \neq w_{ji}$, which would carry information about the distribution of calls between any two connected individuals. Yet, given that there is no *a priori* reason to assume that the individual responsible for initiating the call should interact more strongly (after all, both have exactly the same call duration), we have neglected the directed nature of the links.

We allowed for the possibility that there are some very short calls which, when mapped to links, could affect the overall topology of the network. To see this, we filtered out links with total call duration less than 10 seconds per link over the examined period of 18 weeks. After this we filtered out nodes with strengths less than 60 seconds per node over the period, such that if a node is filtered out, the links connected to it are also removed. These extremely short calls do not in general represent true phone numbers, but rather mobile phone and service updates. Indeed, a common way of obtaining a new handset is by signing up for a new service, and after its activation the users switch back to the old number. We call a network without the reciprocity requirement a *non-mutual network*, i.e. a one-directional call between *A* and *B* is sufficient for them to be linked together. In contrast, a network in which the calls are required to be reciprocal is called a *mutual network*. The results of these filterings are shown in Table 1. It turns out that imposing the reciprocity condition does eliminate some of the outliers, which can be best seen in the degree distribution plots (Fig. S1). However, filtering seems to have little effect on any of the studied distributions. Consequently, in this study we used a mutual network constructed from unfiltered data.

|  | Original | Link filtered | Node filtered |
|---|---|---|---|
| Non-mutual network | | | |
| $N$ | 7190101 | 7087285 | 6742843 |
| $L$ | 22589912 | 21380100 | 21016174 |
| Mutual network | | | |
| $N$ | 4646014 | 4556373 | 4304659 |
| $L$ | 6997180 | 6714614 | 6682364 |

**Table S1.** The number of nodes *N* and links *L* in the original, link filtered, and node filtered networks. Going from non-mutual to mutual network changes the size of the system, both in terms of *N* and *L*, whereas there are very small differences between the original, link filtered, and node filtered networks.

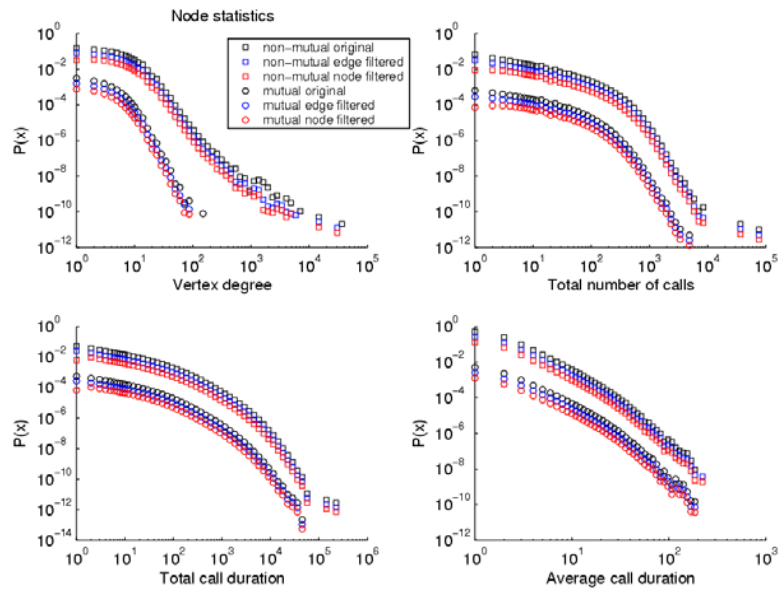

**Fig. S1.** Node strength distributions for mutual and non-mutual networks under different filterings. The curves have been shifted vertically for clarity of presentation. The total number of calls is the sum of the number of calls placed between $v_i$ and $v_j$ in either direction. Similarly, the total call duration is the time $v_i$ and $v_j$ have spent on the phone. Average call duration is the total time spent on the phone divided by the number of calls, both taken over the examined time period.

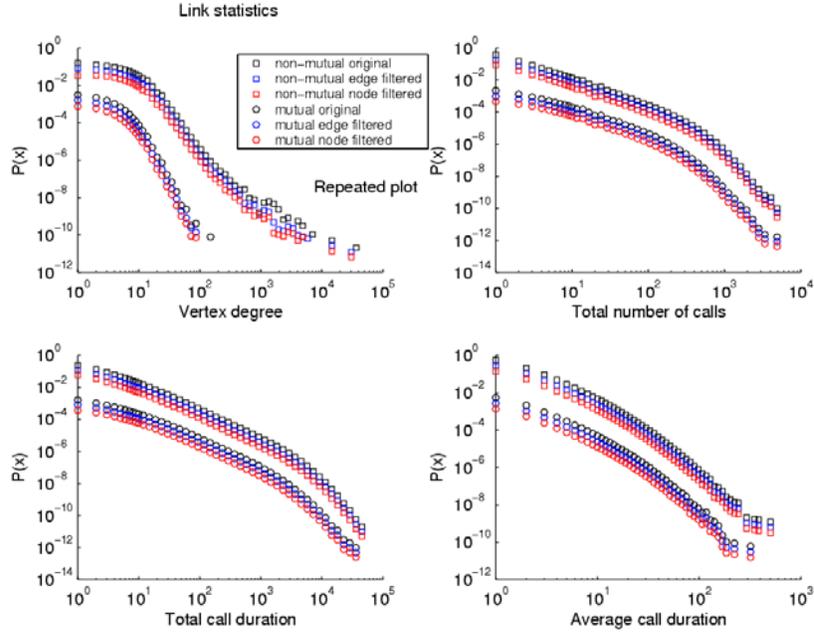

**Fig. S2.** Link weight distributions for mutual and non-mutual networks under different filterings. Note that the top left plot is the same as in the previous figure.

**Weak ties conjecture**

A direct conjecture of the weak ties hypothesis is that communities are locally connected by single weak ties (1). Granovetter justifies this conjecture by framing the hypothesis more precisely in order to derive its implications for larger networks: "The triad which is most unlikely to occur, under the hypothesis stated above, is that in which *A* and *B* are strongly linked, *A* has a strong tie to some friend *C*, but the tie between *C* and *B* is absent." Assuming that this structure never happens, he arrives at the conjecture.

We can obtain the conjecture also using slightly different reasoning. Assume that there is a single tie *A-B*, known as a local bridge, connecting two communities and assume that it is strong. Based on the weak ties hypothesis, we expect the neighborhoods of *A* and *B* (which we assume exist) to overlap. But this means that there is another local bridge, a path of length 2, connecting the nodes and, thus, the two communities. This contradicts our assumption about there being just one strong local bridge and, therefore, the bridge must be a weak tie.

**Sampling**

The mobile phone call records, from which the network is constructed, were obtained from a major mobile operator with a market share of approximately 20% in the target country. Although the dataset covers some seven million users, it is nervertheless a sample of the underlying phone call network that consists of all mobile users in the country. In this section we investigate the possible bias of having a finite sample of the underlying network on the result shown in Fig. 1D, i.e., the increase of overlap $\langle O \rangle_w$ as a function of (cumulative) tie strength. We use the term *sample MCG* to denote the network studied in this paper and *population MCG* to denote the entire mobile phone communication network.

Let $p = 0.20$ denote the 20% market share of the operator in the target country. We assume that the nodes are all identical and that the probability of a node being governed by the operator is independent of the probability of its neighbour being governed by the operator. Given these assumptions, we can interpret $p$ as the probability of a randomly chosen node being governed by the operator and, consequently, its being included in the sample. If we use $N$ to denote the number of nodes in the sample MCG, the expected number of nodes in the population MCG is given by $\hat{N} = N/p = 5N$. Similarly, given the above assumptions, the probability for a link in the population MCG to be included in the sample MCG is $p^2$, whereas the probability for a triangle in the population MCG to be included in the sample MCG is $p^3$. Based on the observed sample, the expected number of links and triangles in the population MCG are, therefore, $\hat{L} = L/p^2 = 25L$ and $\hat{T} = T/p^3 = 125T$, respectively, meaning that we would expect the population MCG to contain 25 times the number of links and 125 times the number of triangles present in the sample MCG.

Since the value of $p$ affects the number of observed nodes, links, and triangles in the sample, it is important to consider how it may affect overlap, defined in the text as $O_{ij} = n_{ij}/[k_i + k_j - n_{ij} - 2]$, where $n_{ij}$ is the number of common neighbors of $v_i$ and $v_j$, i.e., the number of triangles around the link $(v_i, v_j)$, and $k_i$ ($k_j$) denotes the degree of node $v_i$ ($v_j$). Of particular importance is the behavior of overlap averaged over links of a given weight as shown in Fig. 1D, denoted with $\langle O | w^D \rangle \equiv \langle O \rangle_w$, where the superscript in $w^D$ emphasizes that we are using aggregated call durations as link weights. To estimate the effect of $p$ on $\langle O | w^D \rangle$, we generate a *resample* by including in it each node in the LCC (largest connectec component) of the sample MCG with a probability $p$. In this sampling scheme, varying probability $p$ results in different sample sizes, and in the limit of setting $p = 1$ we recover the sample MCG. We consider only the LCC of the resulting resample, since for $p < 1$ the network is likely to become fragmented. The motivation for using this sampling procedure is that it mimics the way in which the sample MCG is obtained from the (unobserved) population MCG.

We chose to use $p = 0.8$, $p = 0.6$, and $p = 0.4$, and extracted three samples for each, resulting in a total of nine different samples with average sample sizes of $\langle N_{LCC, p=0.8} \rangle \approx 2.6 \times 10^6$, $\langle N_{LCC, p=0.6} \rangle \approx 1.4 \times 10^6$, and $\langle N_{LCC, p=0.4} \rangle \approx 0.4 \times 10^6$ corresponding, respectively, to the different values of $p$. Of these the $p = 0.4$ case is most interesting: the LCC of the sample MCG contains about $4.0 \times 10^6$ nodes, roughly 10% of the estimated $35 \times 10^6$ mobile phone users in the country, while using using $p = 0.4$ results in a resample of $\langle N_{LCC, p=0.4} \rangle \approx 0.4 \times 10^6$ nodes, roughly 10% of the nodes in the LCC of the sample MCG. The results are shown in Fig. S3. Although lower values of $p$ result in slightly lower values of $\langle O | w^D \rangle$, its qualitative behavior is fairly insensitive to $p$, and the curves have the same characteristic features as the one in Fig. 1D. Further, examining average overlap as a function of cumulative weight (Fig. S3B) shows that the curves become slightly steeper as $p$ increases.

Consequently, it is safe to assume that the behaviour of $\langle O \rangle_w$ is unaffected by the finite sample. Had we access to the records of all mobile phone users in the country and not just those of a single operator, based on Fig. S3B, we would expect an even more pronounced increasing trend for $\langle O \rangle_w$.

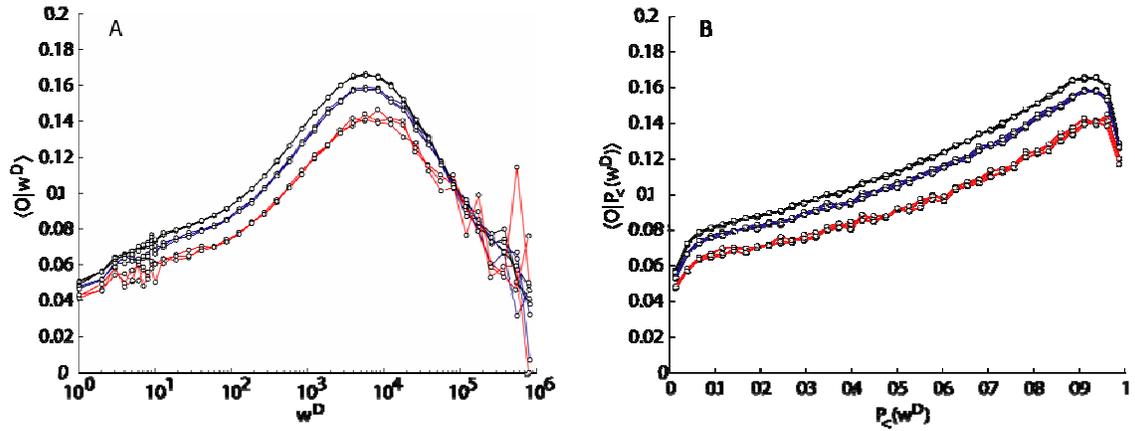

**Fig. S3.** **(A)** Average link overlap $\langle O | w^D \rangle$ as a function of absolute weight, the aggregated call duration $w^D$, and **(B)** average link overlap $\langle O | P_<(w^D) \rangle$ as a function of cumulative weight $P_<(w^D)$, corresponding to the fraction of links with weight less than or equal to $w^D$, for different network samples. There are altogether nine curves in each plot, corresponding to three different samples for each of the three chosen values of extraction probability $p = 0.8$ (black), $p = 0.6$ (blue), and $p = 0.4$ (red). The curves corresponding to different samples for a fixed value of $p$ practically coincide. While lower values of $p$ result in slightly lower values for the average overlap, the qualitative behavior of the curves remains unchanged. This demonstrates that the result of Fig. 1D, i.e., the higher the value of $w^D$ the higher the value of overlap $O$ on average, is not sensitive to having a one-operator-sample of the underlying phone network, and it can be reproduced for sub-samples of the original sample (original data).

**Interdependence of weights and topology**

Fig. S4A shows overlap $O_{ij}$ averaged over all links with weight $w$, as a function of weight $w$, indicating that while for small weights ($w < 10^4$) the overlap $\langle O \rangle_w$ increases with $w$ as expected, for large weights ($w > 10^4$) the overlap $\langle O \rangle_w$ actually decreases. This means that in the region above $w \approx 10^4$, the stronger the tie, the smaller the overlap. This surprising decreasing trend can be understood by considering the link weight distribution shown in Fig. S4B, from which we find that only about 5% of the links lie in the $w>10^4$ region. To correct for this uneven weight distribution in the paper, we plot the overlap as a function of the cumulative link weight $P_{cum}(w)$, which is the percentage of links with weight smaller than $w$, and it is shown in Fig. 1D in the paper.

Since the decreasing trend for top 5% of weights concerns some 325 000 links, it cannot possibly be attributed to insufficient statistics. The links in this region correspond to pairs of users who devote more than three hours to each other over the investigated period. Our measurements indicate, however, that they have a common property: These individuals devote the vast majority of their on-air time to a single acquaintance, and the time spent with others is negligible. Consider a link located between vertices $v_i$ and $v_j$ carrying weight $w_{ij}$, and denote the strengths of the adjacent nodes with $s_i$ and $s_j$, respectively, defined as $s_i = \sum_{j, j \in N(v_i)} w_{ij}$, where the sum index $j$ runs over the neighbours of node $i$. The smaller of the strengths is given by $\min(s_i, s_j)$ and the larger by $\max(s_i, s_j)$, unless the strengths are equal. The ratios $\min(s_i, s_j)/w_{ij}$ and $\max(s_i, s_j)/w_{ij}$, shown in Fig. S4C, correspond to the strengths of the nodes measured in units of the link weight $w_{ij}$. For weak links (small $w_{ij}$) both of these values are high, meaning that overall both adjacent nodes spend a considerably longer time on the phone than they do talking to each other and, thus, the link connecting them constitutes only a small fraction of their on-air time. As we move towards strong links (high $w_{ij}$), we find both ratios decreasing and eventually converging to one at approximately $w=10^4$. This demonstrates that for strong links, in the region where $\langle O \rangle_w$ start to decrease in Fig. S4A, the strengths of both adjacent nodes are about as large as the link weight $w_{ij}$ and, thus, the high weight relationship clearly dominates the on-air time of both users. Consequently, both have less time to interact with other acquaintances, explaining the onset of the decreasing trend for $\langle O \rangle_w$ in Fig. S4A.

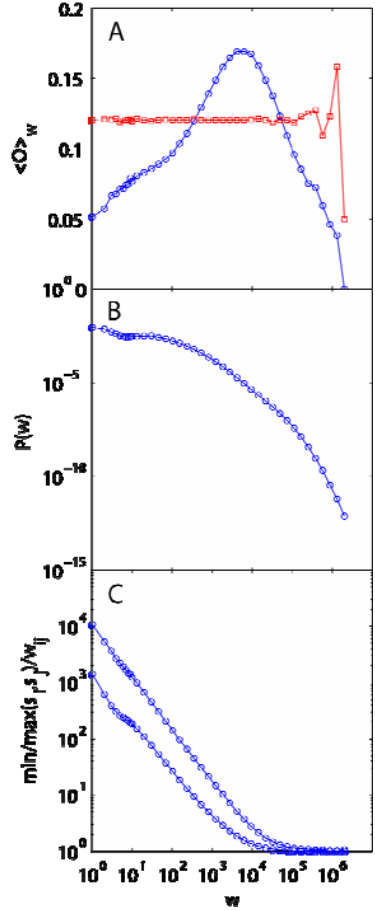

**Fig. S4.** **(A)** The overlap of link neighborhood $\langle O \rangle_w$ increases as a function of the link weight $w$ (blue circles) up-to $w \approx 10^4$, revealing a statistical connection between local network topology and link weights. A random reference (red squares) is obtained by randomly permuting the weights, thus removing the coupling between $\langle O \rangle$ and $w$. Surprisingly, for large weights $w \approx 10^4$ the overlap $\langle O \rangle_w$ actually decreases in this region, apparently contradicting the weak ties hypothesis. Yet, as we explain, that region represents a minority of the users. **(B)** The distribution of links weights $w_{ij}$ decays fast, with only 4.4% mass to the right of $w = 10^4$. This means that the decreasing part of the $O_{ij}$ curve applies to less than 5% of links, which is seen clearly by plotting the overlap $O_{ij}$ as a function of cumulative weight $P_{\text{cum}}(w)$ as in Fig. 1d. **(C)** The fraction of total time (node strength) devoted by the adjacent nodes to a link of weight $w_{ij}$ is given by $\min(s_i, s_j)/w_{ij}$ and $\max(s_i, s_j)/w_{ij}$, and is here plotted as a function of weight $w$. Values close to one indicate that the communication is almost entirely focused on one individual in the $w \approx 10^4$ region.

**Betweenness centrality for links**

For a link $e = (v_i, v_j)$ we can write betweenness centrality $b_{ij}$ as

$$b_{ij} \equiv \sum_{v \in V_s} \sum_{w \in V/\{v\}} \frac{\sigma_{vw}(e)}{\sigma_{vw}} \qquad (2)$$

where $\sigma_{vw}(e)$ is the number of shortest paths between $v_v$ and $v_w$ that contain $e$, and $\sigma_{vw}$ is the total number of shortest paths between $v_v$ and $v_w$ (2). In practice, we use the algorithm introduced in (3) to compute $b_{ij}$ but, due to limited computing capacity, instead of using all the nodes of the set $V$ making up the network, we use a subset $10^5$ nodes in the sample $v_v \in V_s$ as starting points. The size of the set $V_s$ is given by $N_s$.

**Determining the nature and the position of the phase transition point**

The transitions observed in Fig. 3 suggest two important questions: How does the position of the critical threshold $p_c$ depend on the size of the system? Are the transitions genuine phase transitions or finite size effects? In order to answer these questions, we carried out finite size scaling (FSS) for all four different thresholding schemes (remove min $w_{ij}$, min $O_{ij}$, max $w_{ij}$ and max $O_{ij}$ links).

In many large random systems studied by statistical mechanics, from gases to magnetic materials, the system is considered infinite in the number of its constituent elements. Different quantities of interest can be expressed in terms of the correlation (connectivity) length $\xi$ of the system, which in the vicinity of a phase transition diverges like $\xi \sim |p-p_c|^{-\nu}$, where $\nu$ is the critical exponent for correlation length. However, in a finite system, the correlation length is limited by the system size, and the divergence becomes rounded. Consequently, other quantities related to the correlation length also show a rounded signature of the divergence, but never actually diverge due to finite $N$ (4), as demonstrated for path length in Fig. S5. In general, the location of the transition $p_c$ depends on the system size as

$$| p_c(N) - p_c(\infty) | \sim N^{-\chi}, \qquad (3)$$

where $p_c(\infty)$ corresponds to the extrapolated value in the thermodynamic limit as $N \to \infty$. Here the value of the exponent $\chi$ is related to $\nu$ and it quantifies how changing the system size affects the position of the critical threshold, whereas the extrapolated value $p_c(\infty)$ reveals the nature of the transition: If $0 < p_c(\infty) < 1$ there is a real phase transition but, on the other hand, if $p_c(\infty) = 1$ there is no actual phase transition, and the observed signature is caused by the finiteness of the system (5). We use two different sampling techniques to produce systems of different size *N*. In the first approach, we choose randomly one node in the initial network as a source node $v_s$ and extract a radius $\ell$ neighborhood of this node, including all nodes, and links between them, which are at most a distance $\ell$ from $v_s$ (less than $\ell$ links from $v_s$). The size of the sample depends exponentially on the extraction depth $\ell$, so that increasing the value of $\ell$ enables us to extract larger samples, although the realized sample size *N* will depend on the starting node $v_s$ due to the non-homogeneous topology of the network. We call such a sample an *extract*. In the second approach, we generate a sample of the original network by mimicking the process responsible for generating the network. Consider the original network to represent the true underlying social network of which we see only a part, since phone calls are just one form of social interaction. We then assign an occupation probability *p* to each node in the network, corresponding to the probability that this node is governed by the home operator and, thus, belongs to our sample. This means that the probability for a given node to belong to the sample, i.e. to be occupied, is an independent random trial and does not depend on whether its neighbors are occupied. In this sampling scheme varying the node occupation probability *p* results in different sample sizes, just as varying extraction depth $\ell$ does in the extract sampling scheme. The original network corresponds to *p*=1, whereas if *p*<1 the network is likely to become fragmented, in which case we take the largest connected component (LCC) as our sample. A sample obtained using this second method is called a *resample*.

To carry out FSS we need to know the size of the system *N* and the location of the transition $p_c(N)$ for this finite system. Having several $N, p_c(N)$ point pairs, corresponding to different system sizes, allows us to extrapolate the value of $p_c(\infty)$. The system size *N* is just the number of nodes in the given sample and can be obtained easily, but finding the value of $p_c(N)$ is a bit more laborious in practice. In principle, we can find this from the behavior of susceptibility, defined as $S = \sum_s n_s s^2 / \sum_s n_s s$, where $n_s$ is the number of clusters, per lattice site, containing *s* sites and the LCC is excluded from the sum, but in practice we measure $\sum_s n_s s^2$, which behaves similarly to *S*. Although *S* diverges for *N*→∞ at the transition point, in practice it is rather noisy even for medium size systems, making it difficult to pinpoint the location precisely. A more robust technique is to use the smoother, monotonically decreasing order parameter, defined as the fraction of nodes in the LCC and written as $R_{LCC} = \sum_s n_s s$. $R_{LCC}$ is expected to vary most rapidly at the threshold - in fact $\partial R_{LCC} / \partial f$ usually diverges in an infinite system. We can find the location of the transition by computing this derivative numerically and by identifying its steepest descend point with the transition point, which should coincide with the point of divergence for susceptibility. This method works better, but the numerical derivative is not sufficiently robust for smaller systems.

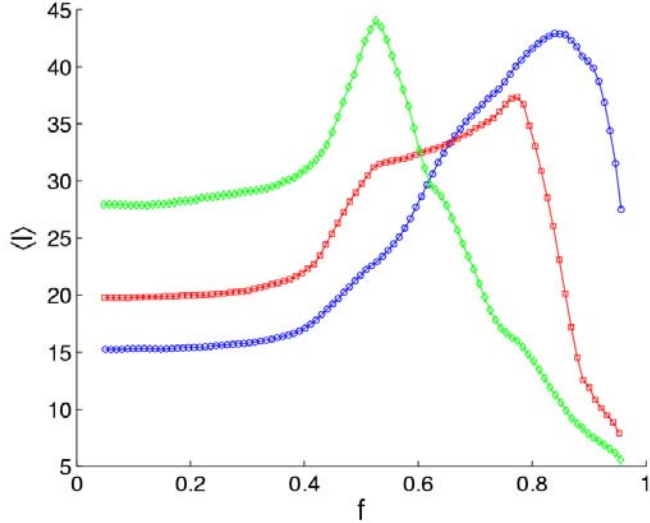

**Fig. S5.** Rounded signature of divergence of average shortest path length $\langle \ell \rangle$ due to finite system size. The green (◊), red ( ), and blue (o) curves are associated with system sizes $N \approx 4.4 \times 10^5$, $N \approx 1.4 \times 10^6$, and $N \approx 3.3 \times 10^6$, and they were obtained using resampling extraction with node occupation probabilities $p=0.40$, $p=0.60$, and $p=0.90$, respectively. For each value of node occupation probability $p$ we sampled a few systems, carried out the thresholding for each of them, computed the average $\langle \ell \rangle$, and then finally smoothened the plot with a moving window average. The critical point $f_c$ moves to the right as the size of the system increases and, since there is no phase transition in this case, we have $f_c \to 1$ as $N \to \infty$. Note that the starting value of $\langle \ell(f=0) \rangle$ is highest for the smallest system, because the networks are more tree like there as relatively speaking more links are missing from small than large samples, suppressing the small world effect of short paths.

Fortunately, Eq. 3 is valid for every reasonable definition of a percolation threshold for finite large systems, and it is only the proportionality constant that is different for different definitions of the onset of percolation (5). It turns out that in this case the most reliable results are obtained by manually determining the transition point $p_c(N)$, denoted in the text with $f_c(N)$, from plots of order parameter $R_{LCC}$ vs. the control parameter $f$. We then make a of plot $f_c(N)$ vs. $1/N$ and fit, in the sense of least sum of squared error, a second order polynomial to the data. The transition point in the infinite size limit is extrapolated from the $y$-intercept of the fit. In some cases the coefficient of the second order term is close to zero, so that the fit effectively is linear. In most cases, however, the fit is clearly curved, indicating that the exponent $x$ is different from -1. Theoretically, the inclusion of the second order term can be justified as a correction to the leading scaling behavior. The correction vanishes as $f \to f_c$, but its contribution may be significant even if $|f - f_c|$ is small. We use this method to obtain sets of estimates of $f_c(\infty)$ for different thresholding schemes and sampling techniques using bootstrapping, in which we randomly choose half of the points to be included in the bootstrap sample, and find out the value of $f_c(\infty)$ using only the points in the bootstrap sample. Repeating this 10000 times gives a distribution of the estimates of $f_c(\infty)$. We take the value $f_c(\infty)$ as the mean of the bootstrap distribution, and the error bounds are taken as the standard deviations of the $f_c(\infty)$ distribution (6).

The results of finite size scaling are given in Table 2. From the extrapolated values it is clear that we have phase transitions as $f_c(\infty) \neq 1$ for descending thresholding, and the values are different for weight-driven and overlap-driven thresholding schemes. The intervals of plausible $f_c(\infty)$ values, taken as the region that is one standard deviation from the mean, are slightly different for extracts and resamples, but part of these intervals coincide. Both sampling techniques support the idea that $f_c(\infty)$ for descending weight thresholding lies in [0.83,0.84], while $f_c(\infty)$ for descending overlap thresholding lies in [0.66,0.67]. Put together, these result fully support the existence of a phase transition for these thresholding schemes.

The results for ascending thresholding are not quite as clear. The coincidence intervals of extracts and resamples for weight and overlap thresholding are [0.89,0.92] and [0.93,0.97], respectively. In the latter case of ascending overlap thresholding, $f_c(\infty) = 1$ is contained within one std of the mean for resamples and within two stds of the mean for extracts, suggesting that instead of a phase transition we most likely have a finite size effect. For ascending weight-driven thresholding $f_c(\infty) = 1$ is not included within two standard deviations of the mean. However, there are two important practical aspects to be kept in mind when interpreting the results for ascending thresholding. First, the behavior of the order parameter $R_{LCC}$ as a function of the control parameter $f$ is noisier for ascending than descending thresholding, with the effect that estimating the finite thresholds $f_c(N)$ is more prone to errors in the ascending scheme. This is a consequence of the structural properties of the studied network. Second, the manual estimates of $f_c(N)$ may have a slight downward bias. Since $f_c(N)$ must lie in the [0,1] interval, one would not estimate $f_c(N) > 1$ for any sample as this does not have any physical meaning. Thus, we conclude that the ascending overlap-driven thresholding exhibits no phase transition; In the case of ascending weight-driven thresholding the transition point is dramatically shifted upward, and is compatible with the assumption of no transition for $f < 1$.

Overall, the network's response to removing weak links is qualitatively different from the response to removing strong links, but quite independent whether we use weights $w_{ij}$ or overlap $O_{ij}$. Our results suggest that the transition observed for removing strong links first is a finite size effect ($f_c = 1$), whereas the transition for removing weak links first is a genuine phase transition ($f_c \neq 1$). This means that the observed qualitative difference between weak and strong links is not a consequence of using the given, finite size sample, but demonstrates that weak and strong links are qualitatively different regardless of the size of the system.

| Scheme | $n$ | $f_c(\infty)$ |
|---|---|---|
| DW (extract) | 17 | 0.80±0.04 |
| DW (resample) | 21 | 0.85±0.02 |
| DO (extract) | 17 | 0.62±0.05 |
| DO (resample) | 21 | 0.69±0.03 |
| AW (extract) | 17 | 0.89±0.03 |
| AW (resample) | 21 | 0.91±0.02 |
| AO (extract) | 17 | 0.92±0.05 |
| AO (resample) | 21 | 0.98±0.05 |

**Table S2.** A summary of finite size scaling results. The key to the different thresholding schemes is the following: A = ascending (insert min links first ≡ remove max links first), D = descending (insert max links first ≡ remove min links first), W = weight driven thresholding, and O = overlap driven thresholding. The words extract and resample in parentheses refer to extracted and resampled samples, respectively, on which the FSS is based. The number of available samples, after the smallest ones were discarded, is denoted with $n$. The number of samples used in each bootstrap realization is $n/2$, and the $f_c(\infty)$ is the value of the percolation threshold extrapolated in the thermodynamic limit as $N \to \infty$.

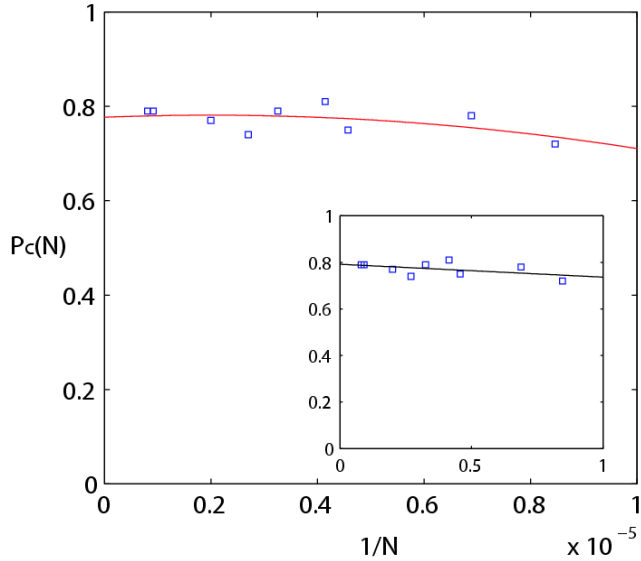

**Fig. S6.** Main panel: One realization of a bootstrap sample for descending weight-driven thresholding, using an extract sample, and the corresponding second order polynomial fit to it. Inset: A linear fit to the same data. Both fits yield practically identical results. The extrapolated value $p_c(\infty)$ for this particular sample is approximately 0.78. Since $p_c(\infty)$ is clearly less than unity, this corresponds to a genuine phase transition. In the case of no transition we would have $p_c(\infty) \approx 1$.

**Spreading model**

Considerations of information flow lead us to formulate a simple model in which the spreading probability from an infected node $v_i$ to its nearest susceptible (non-infected) neighbor $v_j$ was made proportional to the link weight $w_{ij}$. Introducing a constant of proportionality $x$, we write the time independent probability of passing information from $v_i$ to $v_j$ as $P_{ij}=xw_{ij}$, where increasing $x$ results in a higher spreading probability. The most obvious choice is to set $x = 1/\max_{ij}(w_{ij})$, in which case for the globally strongest link we have $P_{ij}=1$, and for all others $P_{ij}<1$. While this is a reasonable choice, it results in extremely long running times for the simulation. The reason for this is the highly skewed weight distribution $P(w)$, so that normalizing with the globally maximum weight, which can be seen as an outlier, results in very low transmission probabilities for most links, requiring a large number of trials before any macroscopic spreading takes place. This problem is amplified by the fact that the simulations, both for empirical and random network, should be carried out for an ensemble.

We can circumvent this problem by lowering the value of $x$, which speeds up the simulations without affecting the qualitative behavior of the system and, thus, it can be seen as a rescaling of the time axis (Fig. 4, A and B). This introduces a cut-off $w^*$ for the transmission probability $P_{ij}$, below which it is linear with respect to $w_{ij}$, and unity for $w_{ij} \geq w^*$ (Fig. S7). But how should one choose the value for $x$ or, alternatively, for the cut-off $w^*$? While the location is to some extent arbitrary, a range of values suggests itself using the following reasoning. For choosing a suitable value for $w^*$, let us deal in terms of the cumulative weight distribution $P_{cum}(w)$, and choose a value for $P_{cum}(w^*)$ instead. The first requirement is that the relationship $P_{ij} \sim w_{ij}$ should be valid for at least half of the links, since otherwise we can hardly say that the two are proportional, and this gives us a lower limit $P_{cum}(w^*) > 0.5$. Since we are interested in the effect of the coupling between weights and topology on a dynamic process, we will stick to a region of link weights in which this observed coupling holds, and from Fig. 1d we see that this is the case up to $P_{cum}(w) \approx 0.95$, giving an upper limit of $P_{cum}(w^*) \leq 0.95$. Within the lower and upper limits, we would like to have as high a value of $P_{cum}(w^*)$ as possible, but also to ensure that we stay away from the region with anomalous behavior (overlap decreasing as a function of weight, a phenomenon that may be specific to the mobile phone network). These heuristics lead us to choose $P_{cum}(w^*) = 0.90$, which for the studied period of 18 weeks corresponds to $w^* = 3867$ seconds, i.e. a little over an hour, or $x = 1/w^* \approx 2.59 \times 10^{-4}$ 1/s. With this choice, the intended relationship $P_{ij} \sim w_{ij}$ holds for 90% of the links, while for the strongest 10% of the links the transmission always takes place. It turns out, however, that the qualitative nature of the spreading results is fairly insensitive to the precise value of $x$, i.e. the weight permuted network performs better at spreading than the empirical network for different values of $x$.

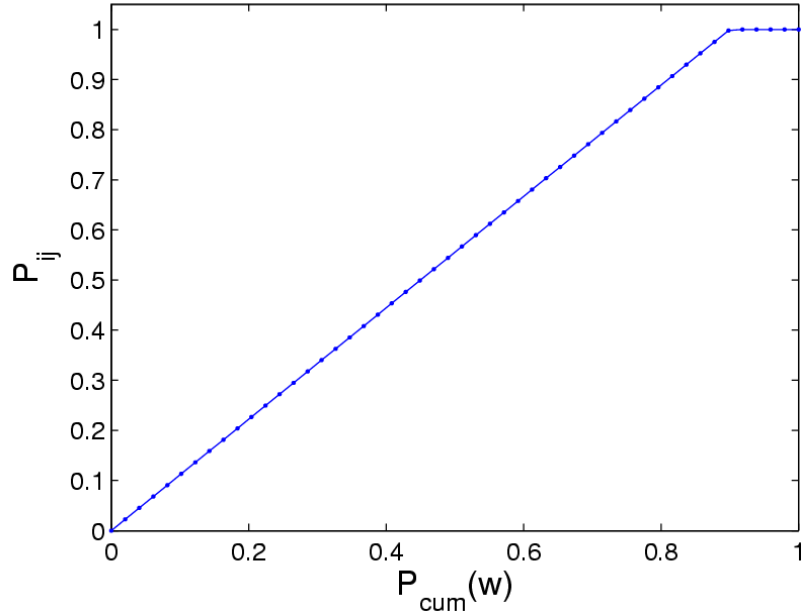

**Fig. S7.** The transfer probability $P_{ij}$ as a function of cumulative link weight $P_{cum}(w)$, the fraction of links with weight less than $w$. Using the value of $x \approx 3.0 \times 10^{-4}$ results in a cut-off at $w^* \approx 0.90$, and thus the intended relationship $P_{ij} \sim w_{ij}$ applies for 90% of links.

Note that although the cut-off was introduced for computational purposes, its existence may, in fact, be a desirable property. Common sense tells us that some pieces of information are more important than others or, in the context of gossip, some pieces of gossip are juicier than others. Lowering the cut-off point $w^*$ means that we have more links with $P_{ij}=1$, such that if $v_i$ has access to information, it will always pass it on to its neighbor $v_j$ as long as $w_{ij} \geq w^*$. For lower cut-off points this will be true for an increasing number of links in the network, and soon rumors will spread like wildfire.